\documentclass[reprint, prd]{revtex4-2}

\usepackage{amsmath,amssymb,amsfonts,bm}
\usepackage{mathtools}

\usepackage{graphicx}
\usepackage{rotating}

\usepackage{listings}

\usepackage{tikz}

\usepackage{CJKutf8}

\usepackage{xspace}
\usepackage{comment}
\usepackage{hyperref} 

\makeatletter
\let\label\ltx@label
\makeatother

\lstdefinestyle{myStyle}{ belowcaptionskip=1\baselineskip, breaklines=true, frame=none, numbers=none, basicstyle=\footnotesize\ttfamily, keywordstyle=\bfseries\color{green!40!black}, commentstyle=\itshape\color{purple!40!black}, identifierstyle=\color{blue}, backgroundcolor=\color{gray!10!white}, }

\DeclareMathOperator{\Tr}{Tr}

\renewcommand{\arraystretch}{1.15}

\definecolor{lime}{HTML}{A6CE39}
\DeclareRobustCommand{\orcidicon}{
	\begin{tikzpicture}
	\draw[lime, fill=lime] (0,0) 
	circle [radius=0.16] 
	node[white] {{\fontfamily{qag}\selectfont \tiny ID}};
	\draw[white, fill=white] (-0.0665,0.095) 
	circle [radius=0.005];
	\end{tikzpicture}
	\hspace{-2mm}
}

\foreach \x in {A, ..., Z}{\expandafter\xdef\csname orcid\x\endcsname{\noexpand\href{https://orcid.org/\csname orcidauthor\x\endcsname}
			{\noexpand\orcidicon}}
   }

\begin{document}
\begin{CJK*}{UTF8}{gbsn}

\title{CHIC: Caley-Hamilton, Invariants and Constants for Neutrino Oscillation Probabilities and Gradients}

\begin{abstract}
We use the Cayley–Hamilton theorem to derive analytic solutions for three-flavor neutrino propagation amplitude in a constant-density medium and their derivative with respect to the mixing parameters. This approach avoids Hamiltonian diagonalization and exploits precomputed matrix invariants to separate the dependence of oscillation probabilities on neutrino energy and propagation baseline. The results are implemented in the CHIC software, which provides simple, fast, and efficient computation of oscillation probabilities and their derivatives. Finally, we demonstrate the value of probability gradients for neutrino data analysis and introduce a complementary visualization, the \emph{oscillograds}, to probe underlying features of neutrino mixing.
\end{abstract}

\author{P.~Fern\'andez-Men\'endez}
\email{pablo.fer@cern.ch}
\affiliation{Donostia International Physics Center DIPC, San Sebasti\'an/Donostia, E-20018, Spain}
\affiliation{MOMA, Instituto Universitario de Ciencias y Tecnologías Espaciales de Asturias (ICTEA), Independencia 13, 33004 Oviedo, Asturias, Spain}
\affiliation{Departamento de Matemáticas, Universidad de Oviedo, Calvo Sotelo 18, 33007 Oviedo, Asturias, Spain}
\affiliation{Universidade de Santiago de Compostela, Instituto Galego de Física de Altas Enerxías (IGFAE), Rúa de Xoaquín Díaz de Rábago, s/n, Santiago de
Compostela, 15705, Spain}

\maketitle
\end{CJK*}

\section{Introduction}\label{sec:intro}
Since the proposal of neutrino oscillations \cite{Pontecorvo:1957cp, 10.1143/PTP.28.870}, the computation of neutrino oscillation probabilities has become a crucial part of understanding and analyzing experimental data and drawing conclusions about the properties of these elusive particles.

After the discovery of neutrino oscillations by Super-Kamiokande \cite{Super-Kamiokande:1998kpq} and SNO \cite{PhysRevLett.87.071301}, and in parallel with the development of more complex neutrino-oscillation experimental programs, several software packages have been developed to compute these transition probabilities in different scenarios and using different approaches \cite{Huber:2007ji, prob3pp, Arg_elles_2022}.

As we enter the precision era of neutrino oscillations with upcoming next-generation experiments such as DUNE \cite{dune_lowexpo}, Hyper-Kamiokande \cite{hk_sensitivity}, and JUNO \cite{thejunocollaboration2024potentialidentifyneutrinomass}, larger exposures and stricter analysis requirements necessitate upgraded tools that compute neutrino-mixing probabilities efficiently and accurately.

The calculation of mixing probabilities tends to be one of the bottlenecks in neutrino data analyses; accordingly, there have recently been several efforts to optimize it \cite{bustamante2019nuoscprobexactgeneralpurposecodecompute, Denton_2024, Arg_elles_2022}. The most relevant is the NuFast software package \cite{Denton_2024}, which provides a very efficient computation and has drastically reduced computation times compared with existing software.

In this work we delve into the optimization of the analytical computation of these probabilities for the three-flavor neutrino scenario with constant-density propagation using the Cayley–Hamilton theorem \cite{caley, hamilton}, and we present a software implementation, CHIC, that provides performance comparable to the NuFast benchmark while offering additional advantages for both computation and insight into neutrino mixing. Furthermore, the proposed implementation is used to efficiently calculate analytical derivatives of the oscillation probabilities with respect to any of their parameters.

In this paper we review the basics of neutrino oscillations in Section~\ref{sec:osc} to set out the problem and to develop the expressions for the oscillation probabilities using the Cayley–Hamilton theorem, diverging from previous studies by focusing on the constants and invariants of the system in Section~\ref{sec:osc_inv}. In Section~\ref{sec:diff} we obtain expressions for the derivatives of the mixing probabilities. Both developments are implemented in the open-access CHIC software package \cite{chic_code}; Section~\ref{sec:app} summarizes the usage and performance. Finally, we explore the benefits of the analytical calculation of probability derivatives in neutrino oscillation data analyses, and showcase the new insights provided by \textit{oscillograds} in Section~\ref{sec:results}, to conclude in Section~\ref{sec:concl}.

\section{Neutrino Oscillations}\label{sec:osc}
Neutrino oscillations arise because the propagation and interaction terms in the Hamiltonian are not simultaneously diagonalizable. Therefore, interaction (flavor) eigenstates are mixtures of the mass eigenstates, and vice versa.

Assuming the standard three-flavor scenario, the two sets of eigenstates are related by a unitary matrix, the PMNS matrix~\cite{10.1143/PTP.28.870}:
\begin{align}\label{eq:pmns}
    U=\begin{pmatrix}
        1 & 0 & 0 \\
        0 & c_{23} & s_{23} \\
        0 & -s_{23} & c_{23} \\
    \end{pmatrix} \cdot
    \begin{pmatrix}
        c_{13} & 0 & s_{13}e^{-i\delta_{CP}} \\
        0 & 1 & 0 \\
        -s_{13}e^{i\delta_{CP}} & 0 & c_{13} \\
    \end{pmatrix} \cdot \nonumber \\
    \begin{pmatrix}
        c_{12} & s_{12} & 0 \\
        {-s_{12}} & c_{12} & 0 \\
        0 & 0 & 1 \\
    \end{pmatrix}
\end{align}

where $s_{ij}=\sin\theta_{ij}$, $c_{ij}=\cos\theta_{ij}$ are the sines and cosines of the mixing angles, and $\delta_{CP}$ is the CP-violating phase (which changes sign for antineutrinos).

At the source, a neutrino is produced in a flavor eigenstate. Assuming an S-wave, along propagation the mass-eigenstate admixture evolves according to the Schrödinger equation,
\begin{equation}\label{eq:sch}
i\frac{\partial}{\partial x}\Psi = H\Psi,
\end{equation}
where $H$ is the $3\times3$ Hamiltonian. In addition to the kinematic term, there is a potential term describing forward elastic scattering with the medium:
\begin{equation}\label{ec:hamiltonian}
H = \frac{1}{2E}T_0 + V,
\end{equation}
with $E$ the neutrino energy. In the flavor basis the kinematic term is
\begin{equation}
T_0 = U\,\mathrm{diag}(0,\Delta m^2_{21},\Delta m^2_{31})\,U^\dagger,
\end{equation}
where $\Delta m^2_{ij}$ are the mass-squared differences. The potential term is diagonal in the flavor basis and accounts for charged-current forward scattering of electron (anti)neutrinos with electrons in the medium:
\begin{equation}
V = \sqrt{2}\,G_F\,n_e\,\mathrm{diag}(1,0,0),
\end{equation}
where $G_F$ is the Fermi constant and $n_e$ the electron number density. This potential has the opposite sign for antineutrinos. In this work we assume constant medium density to obtain analytic expressions relevant for accelerator and reactor experiments; varying-density cases can be approximated from this solution.

The entangled mass-eigenstate superposition collapses again at the interaction point, producing a nonzero probability of flavor change (neutrino oscillations).

When the Hamiltonian is independent of the baseline $x$ (constant density), the oscillation amplitude is solved analytically:
\begin{equation}\label{eq:ampli}
\Psi_{\nu_\alpha\rightarrow\nu_\beta} =
\langle \nu_{\beta} | e^{-i x H} | \nu_{\alpha} \rangle.
\end{equation}

We use the transition probability matrix $P$, whose $\beta\alpha$ element is the transition probability obtained as the component-wise (Hadamard) product of the amplitude matrix and its complex conjugate:
\begin{equation}\label{eq:prob}
P = \Psi \odot \overline{\Psi}
\end{equation}

Additional terms (e.g., from beyond-the-Standard-Model physics) can be included in the Hamiltonian. Although not treated explicitly in what follows, the methods presented can be extended to those cases.

\section{Three-Flavor Neutrino Propagation}\label{sec:osc_inv}

According to Equation~\ref{eq:ampli}, obtaining the neutrino oscillation probabilities reduces to computing the exponential of the Hamiltonian matrix. To that end, we use the Cayley–Hamilton theorem as in \cite{10.1063/1.533270,PhysRevD.111.035003,psf2023008066}, but we focus on invariants of the Hamiltonian to explicitly extract the dependence on the neutrino energy and propagation distance (baseline). This approach shifts the focus from a physics parametrization to the relevant mathematical quantities and provides key points for a highly efficient calculation, as described in Section~\ref{sec:app}.

As customary, we start by shifting the ground energy level of the system by one third of the trace of the Hamiltonian,
\begin{equation}
    \tilde{H} = H - \frac{\Tr(H)}{3}I,
\end{equation}
with $I$ being the identity matrix and 
\begin{equation}
    \Tr(H)=\frac{\Delta m^2_{21} + \Delta m^2_{31}}{2E} + \sqrt{2}\,G_{F}\,n_e
\end{equation}

Physically, this resets the energy scale of the system with respect to its average energy, introducing a global phase that cancels when computing the neutrino oscillation probabilities in Equation~\ref{eq:prob}, {as shown in Equations~\ref{eq:tr_phase} and~\ref{eq:newP}}. Mathematically, dealing with a traceless Hamiltonian simplifies the characteristic polynomial to a reduced cubic polynomial:
\begin{equation}\label{eq:tr_phase}
    \Psi = \exp(-iHx) = e^{-i\frac{\Tr(H)x}{3}} \, \exp(-i\tilde{H}x) = {e^{-i\frac{\Tr(H)x}{3}} \tilde{\Psi}}
\end{equation}
\begin{equation}\label{eq:newP}
    {P = \tilde{\Psi} \odot \overline{\tilde{\Psi}}}
\end{equation}

The Cayley–Hamilton theorem states that any square matrix satisfies its own characteristic polynomial. In our case this implies that the mixing amplitude $\Psi$ can be written as a second-order polynomial in the reduced Hamiltonian, whose coefficients depend solely on its matrix invariants (and thus implicitly on the energy and the baseline): 
\begin{align}\label{eq:CH}
\raggedright
    {\tilde{\Psi}} &= \exp(-i\tilde{H}x) \bmod \left(\tilde{H}^3 {-} \frac{\Tr(\tilde{H}^2)}{2}\tilde{H} - \det(\tilde{H})\right) \nonumber\\&= A(E, x) I + B(E, x) \tilde{H} + C(E, x) \tilde{H}^2 
\end{align}

The functions $A$, $B$ and $C$ can be computed using again the Caley-Hamilton theorem on the eigenvalues of $\tilde{H}$, so that,
\begin{equation}
    \exp(-i\lambda_jx) = A(E, x) + B(E, x)\lambda_j + C(E, x)\lambda_j^2
\end{equation}

To compute the first eigenvalue $\lambda_0$ of $\tilde{H}$ {we use Viète’s trigonometric formula~\cite{vieta}}, which involves $\det(\tilde{H})$ and $\Tr(\tilde{H}^2)$:
\begin{gather}\label{eq:eigenval0}
    \lambda_0 = \sqrt{\frac{2\,Tr(\tilde{H^2})}{3}} \, \cos \left(\frac{\theta_{\tilde{H}}}{3}\right)
\end{gather}

where the angle $\theta_{\tilde{H}}$ is defined as
\begin{equation}\label{eq:theta_vieta}
     \theta_{\tilde{H}} = \arccos \left(
     {\det(\tilde{H})}\sqrt{\cfrac{54}{\Tr(\tilde{H}^2)^3}}\right)\,
\end{equation}

These quantities can be expressed in terms of the invariants of the original Hamiltonian, with an explicit dependence on energy:
\begin{equation}\label{eq:trace_tilde}
    \Tr(\tilde{H}^2) = \Tr(H^2) - \frac{1}{3} \Tr(H)^2
\end{equation}

\begin{equation}\label{eq:det_tilde}
    \det(\tilde{H}) = \det(H) + \frac{1}{6}\Tr(H^2)\Tr(H)
    - \frac{5}{2}\left(\frac{\Tr(H)}{3}\right)^3
\end{equation}

The other two eigenvalues of the reduced Hamiltonian are obtained sequentially as solutions of the resulting quadratic and linear equations in terms of $\lambda_0$: 
\begin{align}\label{eq:eigenval}
    \lambda_1 &= -\frac{\lambda_0}{2} + \sqrt{\frac{\lambda_0^2}{4}-\frac{\det(\tilde{H})}{\lambda_0}}
    \nonumber\\
    \lambda_2 &= -(\lambda_0 +\lambda_1)
\end{align}

Note that the eigenvalues depend only on the neutrino energy, not on the baseline.

Finally, Equation~\ref{eq:CH} can be written explicitly as:
\begin{equation}\label{eq:name_full}
    {\tilde{\Psi}} = \exp(-i\tilde{H}x) = \sum_{j} \mathfrak{J}_{j}(\Vec{\lambda}, x;\tilde{H})
\end{equation}

where $\vec{\lambda}$ is the vector of eigenvalues and the symmetric functions $\mathfrak{J}_{j}$ are
\begin{equation}\label{eq:coefs}
    \mathfrak{J}_{j}(\Vec{\lambda}, x;\tilde{H}) = 
    \dfrac{e^{-i\lambda_j x}}{2\,\lambda_j^2 {+} \pi_j}
    \left(
    \pi_j \, I \, + 
    \lambda_j \, \tilde{H} \, + \tilde{H}^2
    \right) 
\end{equation}

with $\pi_j$ the product of the other two eigenvalues, i.e. $\pi_j=\lambda_i\lambda_k=\det(\tilde{H})/\lambda_j$.

Defining the vector $\vec{\eta}$ with components given by the exponential terms in Equation~\ref{eq:coefs},
\begin{equation}\label{eq:exp}
    \eta_j = \dfrac{e^{-i\lambda_j x}}{2\,\lambda_j^2 {+} \pi_j}
\end{equation}
we obtain a compact expression for the coefficient functions $A(\vec{\lambda},x)$, $B(\vec{\lambda},x)$ and $C(\vec{\lambda},x)$ of Equation~\ref{eq:CH}, where the energy dependence is encoded in the eigenvalues. The amplitude then reads,
\begin{equation}
    {\tilde{\Psi}} = (\Vec{\pi}\cdot\Vec{\eta})\cdot I \, + \, (\Vec{\lambda}\cdot\Vec{\eta})\cdot \tilde{H} \, + \, (\Vec{1}\cdot\Vec{\eta})\cdot \tilde{H}^2
\end{equation}
where $\Vec{\pi}$ is the vector whose components are $\pi_j$ as defined earlier, and the vector $\Vec{\eta}$ carries all the information depending on the propagation baseline. This explicit splitting between eigenvalue computation depending solely on the energy, and the baseline exponential, will be key in the efficient implementation of the CHIC library, Section~\ref{sec:app}, and for Section~\ref{sec:diff} to provide a compact expression for the amplitude's derivative.

\section{Gradient of the Neutrino propagation}\label{sec:diff}

Locally, the variation of the neutrino oscillation probabilities is encoded in the tangent space generated by the partial derivatives of the amplitude with respect to all the parameters entering the Hamiltonian. These derivatives then provide information about how much the mixing probabilities change with a given parameter $\zeta$.

To compute the derivative of the transition amplitude with respect to a given oscillation parameter $\zeta$ entering the Hamiltonian of the system in Equation~\ref{ec:hamiltonian}, we use the integral-derivative expression for a matrix exponential \cite{PhysRev.84.108}. For our case, plugging Equation~\ref{eq:coefs} we obtain the following expression.
\begin{align}\label{eq:dCH}
    \partial_\zeta{\tilde{\Psi}} =&
    -ix\int_0^1 e^{-i\tilde Hxs}\,\, \partial_\zeta\tilde H \, \,e^{-i\tilde Hx(1-s)} \text{d}s =
    \nonumber \\
    \sum_{jj'} ( \pi_j & I+\lambda_j\tilde H + \tilde H^2 ) \partial_\zeta \tilde H 
    ( \pi_{j'}I+\lambda_{j'}\tilde H + \tilde H^2 ) \mathbf{I}_{jj'}
    \end{align}
Being able to rewrite the exponential of the Hamiltonian as a second-order polynomial reduces the expression to an integral over the exponential terms in Equation~\ref{eq:coefs}, encoded in $\mathbf{I}_{jj'}$.
\begin{align}
    \mathbf{I}_{jj} &= \frac{-ix}{(2\,\lambda_j^2 {+} \pi_j)^2} e^{-i\lambda_jx}
    \nonumber\\
    \mathbf{I}_{jj'}=\mathbf{I}_{j'j} &= \frac{1}{\lambda_j-\lambda_{j'}} \cdot 
    \frac{\left( e^{-i\lambda_jx} - e^{-i\lambda_{j'}x}\right)}{(2\,\lambda_j^2 {+} \pi_j)(2\,\lambda_{j'}^2 {+} \pi_{j'})}
\end{align}
Analogously to $\vec{\eta}$, they contain all the dependence on the baseline. With this definition and taking advantage of their symmetry, we can build the following matrix product to compute the relevant coefficients for $\partial_\zeta\Psi$.
\begin{equation}\label{eq:difco}
    \mathcal{S} = (
        \Vec{\pi},
        \Vec{\lambda},
        \Vec{1}
    )^T \,
    \mathbf{I} \,\,
    (
        \Vec{\pi},
        \Vec{\lambda},
        \Vec{1}
    )
\end{equation}
Finally, computing the derivative of the oscillation amplitude is reduced to the derivative of the traceless Hamiltonian, summarized in Appendix~\ref{sec:annex1}. The explicit expression becomes, 
\begin{align}
    \partial_\zeta {\tilde{\Psi}} = &\mathcal{S}_{00}\,\partial_\zeta\tilde{H} +
    \mathcal{S}_{11}\,\tilde{H}\,\partial_\zeta\tilde{H}\,\tilde{H}\ +
    \mathcal{S}_{22}\,\tilde{H}^2\,\partial_\zeta\tilde{H}\,\tilde{H}^2\ + \nonumber\\
    &\mathcal{S}_{01}\,\lbrace\partial_\zeta\tilde{H},\tilde{H}\rbrace +
    \mathcal{S}_{02}\,\lbrace\partial_\zeta\tilde{H},\tilde{H}^2\rbrace + \nonumber\\
    &\mathcal{S}_{12}\,\tilde{H}\,\lbrace \partial_\zeta\tilde{H},\tilde{H}\rbrace\,\tilde{H}\
\end{align}
Here $\{\cdot,\cdot\}$ denotes the anti-commutator, and we have used the symmetry of the matrix $\mathcal{S}$.

The final expression for the derivative of the oscillation probability matrix is given by twice the real part of the component-wise product of the amplitude and its derivative:
\begin{equation}\label{eq:dP}
    \partial_\zeta P = \partial_\zeta {\tilde{\Psi}} \odot \overline{{\tilde{\Psi}}} + {\tilde{\Psi}} \odot \overline{\partial_\zeta {\tilde{\Psi}}} = 2 \, \mathfrak{Re}\left(\partial_\zeta {\tilde{\Psi}} \odot \overline{{\tilde{\Psi}}}\right)
\end{equation}

\section{CHIC Library}\label{sec:app}
The previous results are implemented in the \texttt{C++} CHIC library \cite{chic_code}. The core implementation extracts Hamiltonian constants and invariants to make the energy and baseline dependence explicit, minimizing matrix operations. For the remaining matrix products, the code uses the highly optimized Eigen \texttt{C++} linear-algebra library \cite{eigenweb}.

CHIC provides two classes: a base class, \texttt{CHIC}, which computes oscillation probabilities, and a derived class, \texttt{CHICDIFF}, which additionally computes analytical derivatives of the probabilities. Both classes allow changing any oscillation parameter via public methods and retrieving relevant quantities such as the Hamiltonian or the eigenvalues.

The next example shows a minimal \texttt{C++} program that computes the oscillation probabilities and prints the results.
\begin{lstlisting}[language=C++, style=MyStyle]
#include "CHIC.h"

int main() {
    double energy = 0.7;  // GeV
    double baseline = 295.;  // km

    CHIC chic;
    
    chic.update_dcp(4.1);  // dcp in rad (234 deg)
    chic.update_dm231(2.5e-3); // eV^2/c^4
    chic.update_dm221(7.5e-5); // eV^2/c^4
    chic.update_th12(0.583638); // theta_12 in rad (33.44 deg)
    chic.update_th13(0.149574); // theta_13 in rad (8.57 deg)
    chic.update_th23(0.858701); // theta_23 in rad (49.2 deg)
    chic.update_density(3.0); // g/cm^3
    
    Eigen::Matrix3d prob;
    prob = chic.compute_oscillations(
        energy, baseline);
    return 0;
}
\end{lstlisting}

For derivatives, an additional argument is required: the parameter with respect to which the amplitude is differentiated.
\begin{lstlisting}[language=C++, style=MyStyle]
#include "CHIC.h"

int main() {
    double energy = 0.7;  // GeV
    double baseline = 295.;  // km

    CHICDIFF dchic;
    
    Eigen::Matrix3d prob, dprob;
    prob = dchic.compute_oscillations(
        energy, baseline);
    dprob = dchic.compute_oscillations_derivatives(
        "dcp", energy, baseline);
    dprob = dchic.compute_oscillations_derivatives(
        "density", energy, baseline);
    dprob = dchic.compute_oscillations_derivatives(
        "th23", energy, baseline);
    return 0;
}
\end{lstlisting}

In addition to the C++ library, an optional Python module, \texttt{pychic}, can be compiled using \texttt{pybind11} \cite{pybind11}. Its usage is analogous to the C++ API.
\begin{lstlisting}[language=Python, style=MyStyle]
import pychic

Enu = 2.5 # GeV
L = 1300. # km

# Using CHIC
ch = pychic.CHIC("antineutrino")
P = ch.compute_oscillations(Enu, L)

# Using CHICDIFF
dch = pychic.CHICDIFF("antineutrino")
P = dch.compute_oscillations(Enu, L)
dP = dch.compute_oscillations_derivatives("dm221", Enu, L)
\end{lstlisting}

In both cases neutrino oscillations are assumed, with the ability to switch to antineutrinos when calling the class as shown in the last example. By default the mixing parameters are those from the latest NuFit analysis \cite{Esteban_2024}.

In terms of performance, the code has been tested against the NuFast code on a 64~GB laptop with Intel$^\text{®}$ Xeon(R) CPU E3-1505M v6 @ 3.00~GHz × 8 processor, both compiled with the same optimizations (\texttt{-std=c++17 -O3 -march=native -ffast-math -funroll-loops -DNDEBUG -fPIC}). NuFast yielded consistently a computation time of around 161~ns for all nine mixing probability channels. In turn, the tests with CHIC need to be addressed specifying which quantities vary. Taking advantage of the separation between energy and baseline dependencies, CHIC can cache quantities that depend only on energy and thereby save a significant fraction of the calculations when only the baseline changes. The results are summarized in Table~\ref{tab:performance}.

\begin{table}[h]
    \centering
\begin{tabular}{p{1cm}|c|c|c}
     & NuFast (all) & CHIC (energy) & CHIC (baseline) \\\hline
Time & 161~ns      & 193~ns        & 114~ns
\end{tabular}
\caption{Performance comparison in the computation of the 3-by-3 oscillation probability matrix by NuFast and CHIC, where varying-energy (center) and baseline-varying (right) cases are shown.}
    \label{tab:performance}
\end{table}

Both NuFast and CHIC have comparable computation times. However, CHIC’s ability to cache energy-only quantities yields a significant 30\% reduction in computation time when applicable. This feature is relevant in scenarios with varying baselines (e.g., reactor experiments) or in {future} implementations that approximate propagation through varying-density media {as propagation through layers of constant density (e.g. neutrinos traveling through Earth)}.

\section{Results and Applications}\label{sec:results}

Besides CHIC's performance, the additional feature of providing gradients of the oscillation probabilities eases the exploration of the local structure of those probabilities.

A primary application is neutrino data analysis. As shown in \cite{PhysRevX.13.041055}, when profiling over systematic uncertainties, computing the analytic gradient of the negative log-likelihood boosts minimization, with the reduction in computation time proportional to the number of nuisance parameters. CHIC provides the tools to extend this approach to the mixing parameters without relying on numerical differentiation.

Similarly, derivative information enables precise calculation of oscillation probabilities when interpolating, reducing the number of evaluation points needed. Therefore these partial derivatives provide an efficient way to alleviate the computational cost of numerical methods associated with neutrino mixing probabilities in data analyses.

\begin{figure}[ht!]
    \centering
    \includegraphics[width=0.9\linewidth]{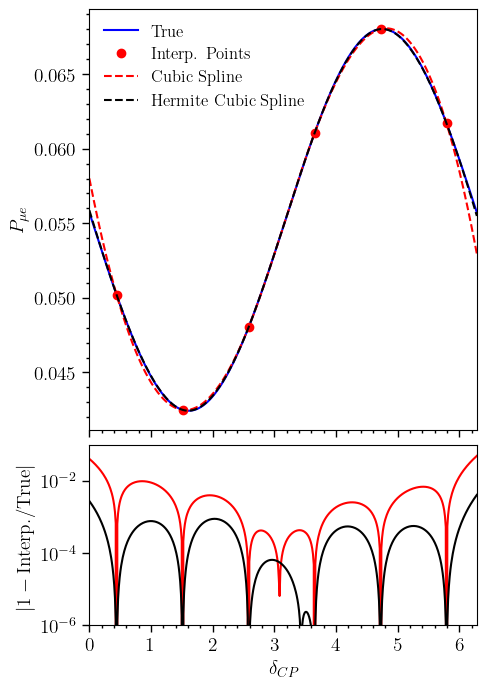}
    \caption{Comparison of cubic spline interpolation with (Hermite) and without the derivative information for the oscillation probability of $\nu_\mu$ to $\nu_e$ assuming an energy of 0.6~GeV and a baseline of 295~km.}
    \label{fig:dcp_interp}
\end{figure}

This feature is illustrated in Figure~\ref{fig:dcp_interp} for the appearance channel in the T2K and Hyper-Kamiokande (T2HK) long-baseline (LBL) setup. There, a Hermite interpolation using only six points matches the true oscillation probability with a relative difference below $10^{-4}$.

Besides providing probabilities, the gradients give information about the local behavior of the probability — not just its value in parameter space. This structure can be studied and visualized by plotting the partial derivative with respect to a given parameter alongside the usual probability plots that scan different parameter values. To that end, we introduce \emph{oscillograds} to probe the underlying features of neutrino mixing.

\begin{figure}[ht!]
    \centering
    \includegraphics[width=0.9\linewidth, height=13.5cm]{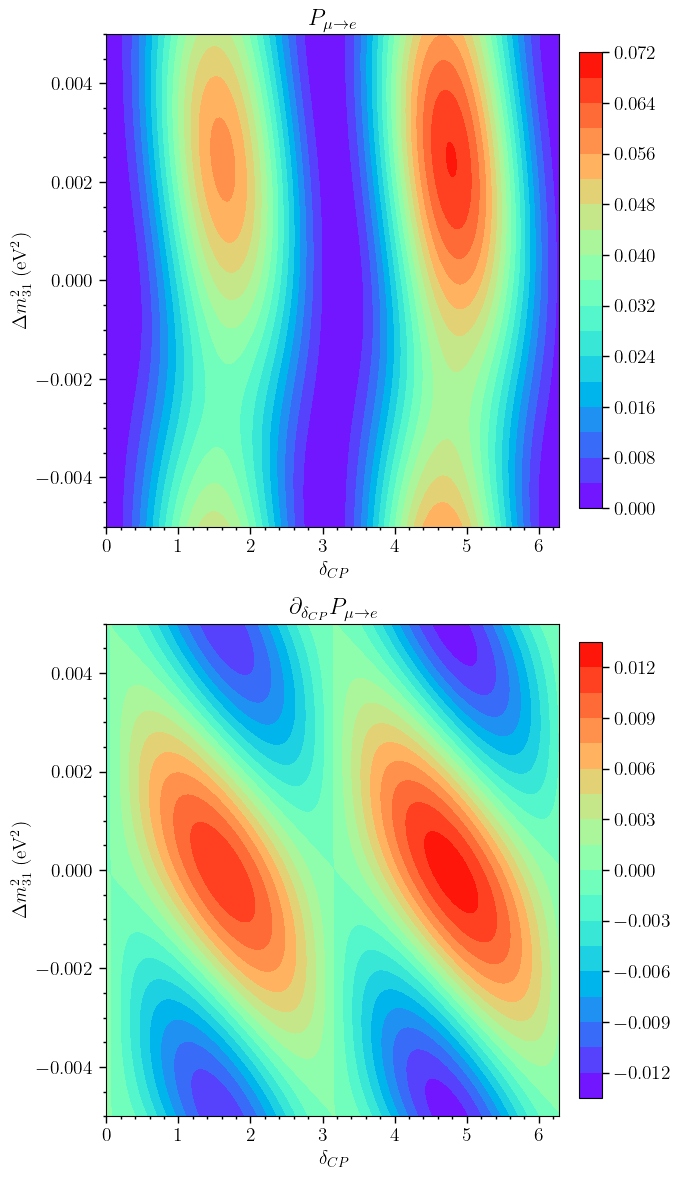}
    \caption{Two-dimensional plots showing the transition probability (oscillogram, top) and its derivative with respect to $\delta_{CP}$ (oscillograd, bottom) of $\nu_\mu$ to $\nu_e$ assuming an energy of 0.6~GeV and a baseline of 295~km, and as a function of $\Delta m^2_{31}$ (vertical axis) and $\delta_{CP}$ (horizontal axis). }
    \label{fig:hk_oscillograds}
\end{figure}

For illustration, Figure~\ref{fig:hk_oscillograds} shows the mixing probability and its derivative with respect to $\delta_{CP}$ for the appearance channel (muon neutrinos to electron neutrinos) as functions of $\Delta m^2_{31}$ and $\delta_{CP}$ for the T2K and T2HK long-baseline (LBL) setups, with neutrino energy 0.6~GeV and baseline 295~km. The top panel shows the quasi-degeneracy of the probability with the mass-squared splitting. The bottom panel shows regions where the rate of change of the probability with $\delta_{CP}$ differs, revealing a slight breaking of the degeneracy between mass ordering and the CP phase around $\pi$. For reference, this plot shows no significant difference between DUNE and T2HK experiments.

\begin{figure}[ht!]
    \centering
    \includegraphics[width=0.85\linewidth, height=13.75cm]{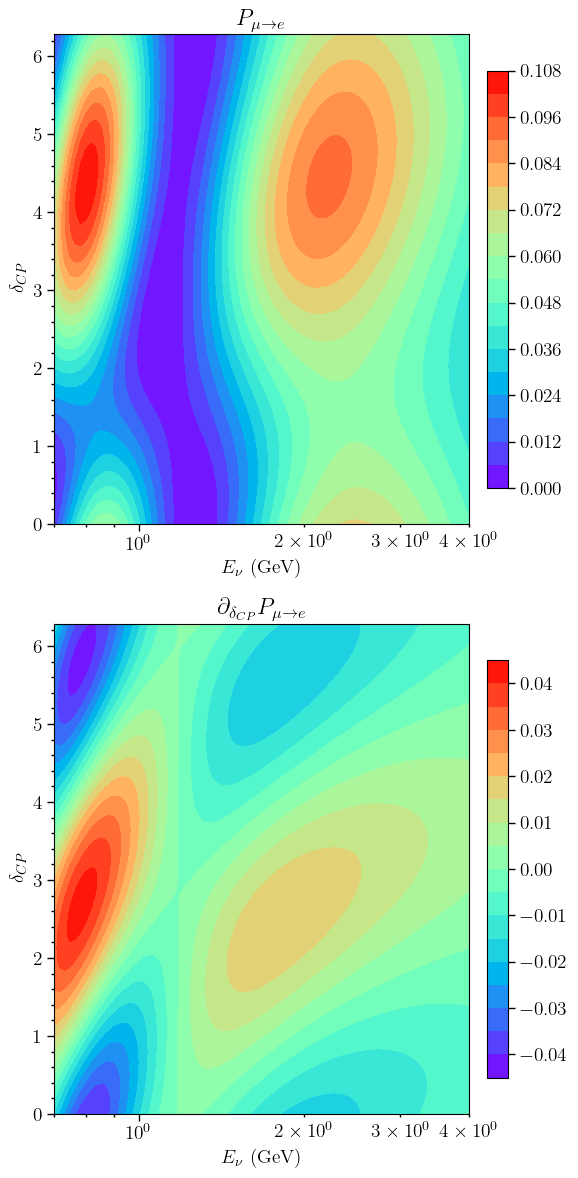}
    \caption{Two-dimensional plots showing the transition probability (oscillogram, top) and its derivative with respect to $\delta_{CP}$ (oscillograd, bottom) of $\nu_\mu$ to $\nu_e$ assuming a baseline of 1,300~km, and {as a function of $\delta_{CP}$ (vertical axis) and energy (horizontal axis).} }
    \label{fig:dcp_dune_oscillograds}
\end{figure}

In this context, we also examine the $\delta_{CP}$ oscillograds for DUNE as a function of energy. DUNE far detector will be exposed to a wide neutrino beam spectrum peaking between 1.5 and 4~GeV. In Figure~\ref{fig:dcp_dune_oscillograds}, the probability for the appearance channel shows two oscillation maxima, the first one at 0.9~GeV and the second one between 1.5 and 2.0~GeV with very little variation across the range of $\delta_{CP}$. However, the oscillograd below shows more clearly how those probabilities vary locally not only with $\delta_{CP}$ but also with energy. These features become relevant when aiming to measure $\delta_{CP}$ at the few percent level and demand a detailed analysis beyond the scope of this work.

Thus, while a sizable mixing probability implies detectability, derivatives show how strongly that probability changes with a given parameter. This information can be interpreted as the achievable precision for that parameter in a given region of parameter space, \cite{hk_sensitivity}. As neutrino physics enters the precision era, such information provides a more complete picture of mixing probabilities for assessing experimental sensitivity.

\section{Conclusions}\label{sec:concl}

In this work we provide the analytic computation of three-flavor neutrino oscillation probabilities for propagation through a constant-density medium using the Cayley–Hamilton theorem. This approach exploits the Hamiltonian structure to separate the energy and baseline dependencies and uses matrix invariants, resulting in a fast and efficient method to compute oscillation probabilities. We also derive analytic expressions for the partial derivatives of the probabilities with respect to the parameters appearing in the Hamiltonian.

Both results are implemented in the CHIC software, which enables fast and efficient computation of oscillation probabilities. The implementation paves the way for efficient software for other scenarios, including varying-density media \cite{Maltoni_2023}, beyond–Standard-Model phenomena, and additional sterile neutrinos.

Finally, given that the analytic derivatives are readily computed by CHIC, we discuss the advantages and extra information these derivatives provide, motivate their use in neutrino data analyses, and analyze how derivatives complement standard probabilities by characterizing the local behavior of mixing probabilities and the experimentally achievable precision.

\section{Acknowledgements}
The author acknowledges helpful inputs from J. P. Dalseno, I. Mart\'inez-Soler, P. Denton, T. Lux and K. Skwarczynski and thanks the support of DIPC and IGFAE. This work has been supported partially by the \textit{Proyectos de Generación de Conocimiento} PID2022-141960NA-I00 and PID2024-162224NA-I00 of the Spanish Ministry of Science, Innovation and Universities.

\bibliography{biblio}

@ARTICLE{vieta,
       author = {{Viète}, F.},
        title = "{‘On the recognition of equations’ and ‘On the emendation of equations’}",
      journal = {Anderson A (Ed.), Opera mathematica},
         year = 1615,
       volume = {III 9},
       country = {France}
}

@ARTICLE{caley,
       author = {{Cayley}, Arthur},
        title = "{A Memoir on the Theory of Matrices}",
      journal = {Philosophical Transactions of the Royal Society of London Series I},
         year = 1858,
        month = jan,
       volume = {148},
        pages = {17-37},
       adsurl = {https://ui.adsabs.harvard.edu/abs/1858RSPT..148...17C}
}

@article{Huber:2007ji,
    author = "Huber, Patrick and Kopp, Joachim and Lindner, Manfred and Rolinec, Mark and Winter, Walter",
    title = "{New features in the simulation of neutrino oscillation experiments with GLoBES 3.0: General Long Baseline Experiment Simulator}",
    eprint = "hep-ph/0701187",
    archivePrefix = "arXiv",
    reportNumber = "TUM-HEP-656-07",
    doi = "10.1016/j.cpc.2007.05.004",
    journal = "Comput. Phys. Commun.",
    volume = "177",
    pages = "432--438",
    year = "2007"
}

@article{hamilton,
    author = {W. Hamilton},
    title = {On the Existence of a Symbolic and Biquadratic Equation which is satisfied by the Symbol of Linear or Distributive Operation on a Quaternion},
    journal = {The London, Edinburgh, and Dublin Philosophical Magazine and Journal of Science},
    ISSN = {1478-6435},
number = {iv. 24: 127–128},
    year = 1862
}

@article{PhysRevX.13.041055,
  title = {Measuring Oscillations with a Million Atmospheric Neutrinos},
  author = {Arg\"uelles, C. A. and Fern\'andez, P. and Mart\'{\i}nez-Soler, I. and Jin, M.},
  journal = {Phys. Rev. X},
  volume = {13},
  issue = {4},
  pages = {041055},
  numpages = {31},
  year = {2023},
  month = {Dec},
  publisher = {American Physical Society},
  doi = {10.1103/PhysRevX.13.041055},
  url = {https://link.aps.org/doi/10.1103/PhysRevX.13.041055}
}

@article{Denton_2024,
   title={Fast and accurate algorithm for calculating long-baseline neutrino oscillation probabilities with matter effects},
   volume={110},
   ISSN={2470-0029},
   url={http://dx.doi.org/10.1103/PhysRevD.110.073005},
   DOI={10.1103/physrevd.110.073005},
   number={7},
   journal={Physical Review D},
   publisher={American Physical Society (APS)},
   author={Denton, Peter B. and Parke, Stephen J.},
   year={2024},
   month=oct }

@article{Maltoni_2023,
   title={From ray to spray: augmenting amplitudes and taming fast oscillations in fully numerical neutrino codes},
   volume={2023},
   ISSN={1029-8479},
   url={http://dx.doi.org/10.1007/JHEP11(2023)033},
   DOI={10.1007/jhep11(2023)033},
   number={11},
   journal={Journal of High Energy Physics},
   publisher={Springer Science and Business Media LLC},
   author={Maltoni, Michele},
   year={2023},
   month=nov }

@article{Pontecorvo:1957cp,
    author = "Pontecorvo, B.",
    title = "{Mesonium and anti-mesonium}",
    journal = "Sov. Phys. JETP",
    volume = "6",
    pages = "429",
    year = "1957"
}

@article{Esteban_2024,
   title={NuFit-6.0: updated global analysis of three-flavor neutrino oscillations},
   volume={2024},
   ISSN={1029-8479},
   url={http://dx.doi.org/10.1007/JHEP12(2024)216},
   DOI={10.1007/jhep12(2024)216},
   number={12},
   journal={Journal of High Energy Physics},
   publisher={Springer Science and Business Media LLC},
   author={Esteban, Ivan and Gonzalez-Garcia, M. C. and Maltoni, Michele and Martinez-Soler, Ivan and Pinheiro, João Paulo and Schwetz, Thomas},
   year={2024},
   month=dec }

@misc{pybind11,
   author = {Wenzel Jakob and Jason Rhinelander and Dean Moldovan},
   year = {2016},
   note = {https://github.com/pybind/pybind11},
   title = {pybind11 — Seamless operability between C++11 and Python}
}

@misc{prob3pp,
   author = {Roger A. Wendell},
   year = {2018},
   note = {https://github.com/rogerwendell/Prob3plusplus},
   title = {Prob3++}
}

@MISC{eigenweb,
  author = {Ga\"{e}l Guennebaud and Beno\^{i}t Jacob and others},
  title = {Eigen},
  howpublished = {https://libeigen.gitlab.io},
  year = {2010}
 }

@misc{chic_code,
  author = {P. Fernández},
  title = {CHIC},
  year = {2025},
  publisher = {GitHub},
  howpublished = {\url{https://github.com/pabloferm/CHIC/releases/tag/v1.0.1}},
}

@article{PhysRev.84.108,
  title = {An Operator Calculus Having Applications in Quantum Electrodynamics},
  author = {Feynman, Richard P.},
  journal = {Phys. Rev.},
  volume = {84},
  issue = {1},
  pages = {108--128},
  numpages = {0},
  year = {1951},
  month = {Oct},
  publisher = {American Physical Society},
  doi = {10.1103/PhysRev.84.108},
  url = {https://link.aps.org/doi/10.1103/PhysRev.84.108}
}

@article{10.1063/1.533270,
    author = {Ohlsson, Tommy and Snellman, H\r{a}kan},
    title = {Three flavor neutrino oscillations in matter},
    journal = {Journal of Mathematical Physics},
    volume = {41},
    number = {5},
    pages = {2768-2788},
    year = {2000},
    month = {05},
    issn = {0022-2488},
    doi = {10.1063/1.533270},
    url = {https://doi.org/10.1063/1.533270},
}

@article{PhysRevD.111.035003,
  title = {Revisiting series expansions of neutrino oscillation and decay probabilities in matter},
  author = {Gr\"onroos, Jesper and Ohlsson, Tommy and Vihonen, Sampsa},
  journal = {Phys. Rev. D},
  volume = {111},
  issue = {3},
  pages = {035003},
  numpages = {17},
  year = {2025},
  month = {Feb},
  publisher = {American Physical Society},
  doi = {10.1103/PhysRevD.111.035003},
  url = {https://link.aps.org/doi/10.1103/PhysRevD.111.035003}
}

@Article{psf2023008066,
AUTHOR = {Chattopadhyay, Dibya S. and Chakraborty, Kaustav and Dighe, Amol and Goswami, Srubabati},
TITLE = {Oscillation and Decay of Neutrinos in Matter: An Analytic Treatment},
JOURNAL = {Physical Sciences Forum},
VOLUME = {8},
YEAR = {2023},
NUMBER = {1},
ARTICLE-NUMBER = {66},
URL = {https://www.mdpi.com/2673-9984/8/1/66},
ISSN = {2673-9984},
DOI = {10.3390/psf2023008066}
}

@article{10.1143/PTP.28.870,
    author = {Maki, Ziro and Nakagawa, Masami and Sakata, Shoichi},
    title = {Remarks on the Unified Model of Elementary Particles},
    journal = {Progress of Theoretical Physics},
    volume = {28},
    number = {5},
    pages = {870-880},
    year = {1962},
    month = {11},
    issn = {0033-068X},
    doi = {10.1143/PTP.28.870},
    url = {https://doi.org/10.1143/PTP.28.870},
    eprint = {https://academic.oup.com/ptp/article-pdf/28/5/870/5258750/28-5-870.pdf},
}

@article{Arg_elles_2022,
   title={nuSQuIDS: A toolbox for neutrino propagation},
   volume={277},
   ISSN={0010-4655},
   url={http://dx.doi.org/10.1016/j.cpc.2022.108346},
   DOI={10.1016/j.cpc.2022.108346},
   journal={Computer Physics Communications},
   publisher={Elsevier BV},
   author={Argüelles, Carlos A. and Salvado, Jordi and Weaver, Christopher N.},
   year={2022},
   month=aug, pages={108346} }

@ARTICLE{dune_lowexpo,
  title = {Low exposure long-baseline neutrino oscillation sensitivity of the DUNE experiment},
  author = {Abud, A. Abed and others},
  collaboration = {DUNE Collaboration},
  journal = {Phys. Rev. D},
  volume = {105},
  issue = {7},
  pages = {072006},
  numpages = {32},
  year = {2022},
  month = {Apr},
  publisher = {American Physical Society},
  doi = {10.1103/PhysRevD.105.072006},
  url = {https://link.aps.org/doi/10.1103/PhysRevD.105.072006}
}

@article{PhysRevLett.87.071301,
  title = {{Measurement of the Rate of ${\ensuremath{\nu}}_{e}+\mathit{d}\ensuremath{\rightarrow}\mathit{p}+\mathit{p}+{\mathit{e}}^{\ensuremath{-}}$ Interactions Produced by $^{8}B$ Solar Neutrinos at the Sudbury Neutrino Observatory}},
  author = {Ahmad, Q. R. and others},
  collaboration = {SNO Collaboration},
  journal = {PRL},
  volume = {87},
  issue = {7},
  pages = {071301},
  numpages = {6},
  year = {2001},
  month = {Jul},
  publisher = {American Physical Society},
  doi = {10.1103/PhysRevLett.87.071301},
  url = {https://link.aps.org/doi/10.1103/PhysRevLett.87.071301}
}

@ARTICLE{Super-Kamiokande:1998kpq,
    author = "Fukuda, Y. and others",
    collaboration = "Super-Kamiokande",
    title = "{Evidence for oscillation of atmospheric neutrinos}",
    eprint = "hep-ex/9807003",
    archivePrefix = "arXiv",
    reportNumber = "BU-98-17, ICRR-REPORT-422-98-18, UCI-98-8, KEK-PREPRINT-98-95, LSU-HEPA-5-98, UMD-98-003, SBHEP-98-5, TKU-PAP-98-06, TIT-HPE-98-09",
    doi = "10.1103/PhysRevLett.81.1562",
    journal = "Phys. Rev. Lett.",
    volume = "81",
    pages = "1562--1567",
    year = "1998"
}

@misc{thejunocollaboration2024potentialidentifyneutrinomass,
      title={Potential to Identify the Neutrino Mass Ordering with Reactor Antineutrinos in JUNO}, 
      author={The JUNO Collaboration},
      year={2024},
      eprint={2405.18008},
      archivePrefix={arXiv},
      primaryClass={hep-ex},
      url={https://arxiv.org/abs/2405.18008}, 
}

@misc{hk_sensitivity,
      title={Sensitivity of the Hyper-Kamiokande experiment to neutrino oscillation parameters using acceleration neutrinos}, 
      author={Kamiokande Collaboration},
      year={2025},
      eprint={2505.15019},
      archivePrefix={arXiv},
      primaryClass={hep-ex},
      url={https://arxiv.org/abs/2505.15019}, 
}

@misc{bustamante2019nuoscprobexactgeneralpurposecodecompute,
      title={NuOscProbExact: a general-purpose code to compute exact two-flavor and three-flavor neutrino oscillation probabilities}, 
      author={Mauricio Bustamante},
      year={2019},
      eprint={1904.12391},
      archivePrefix={arXiv},
      primaryClass={hep-ph},
      url={https://arxiv.org/abs/1904.12391}, 
}


\clearpage
\newpage

\onecolumngrid
\appendix

\ifx \standalonesupplemental\undefined
\setcounter{page}{1}

\setcounter{figure}{0}

\setcounter{table}{0}
\setcounter{equation}{0}
\fi
\renewcommand{\thepage}{Supplemental Materials -- S\arabic{page}}

\renewcommand{\figurename}{SUPPL. FIG.}

\renewcommand{\tablename}{SUPPL. TABLE}
\renewcommand{\theequation}{A\arabic{equation}}

\newcounter{SIfig}
\renewcommand{\theSIfig}{SUPPL. FIG. \arabic{SIfig} }

\section{Derivatives of the Hamiltonian}\label{sec:annex1}
\subsection{Electron Density, \texorpdfstring{$n_e$}{ne}}

\begin{equation}
    \partial_{n_e} \tilde{H} = \sqrt{2}\, G_F \cdot \text{diag}\left( 1, 0, 0 \right) - \frac{\sqrt{2}\, G_F}{3} I
\end{equation}

\subsection{\texorpdfstring{$\Delta m^2_{21}$}{dm21}}

\begin{equation}
    \partial_{\Delta m^2_{21}} \tilde{H} = \frac{1}{2E}\,
    U \cdot \begin{pmatrix}
0 & 0 & 0 \\
0 & 1 & 0 \\
0 & 0 & 0
\end{pmatrix}
\cdot U^\dagger
- \frac{1}{6E} I
\end{equation}

\subsection{\texorpdfstring{$\Delta m^2_{31}$}{dm31}}

\begin{equation}
    \partial_{\Delta m^2_{31}} \tilde{H} = \frac{1}{2E}\,
    U \cdot \begin{pmatrix}
0 & 0 & 0 \\
0 & 0 & 0 \\
0 & 0 & 1
\end{pmatrix}
\cdot U^\dagger
- \frac{1}{6E} I
\end{equation}

\subsection{\texorpdfstring{$\theta_{23}$}{th23}}

\begin{equation}
\partial_{\theta_{23}}U = \begin{pmatrix}
U_{20} & U_{21} & U_{22} \\
-U_{10} & -U_{11} & -U_{12} \\
-U_{00} & -U_{01} & -U_{02}
\end{pmatrix} \nonumber
\end{equation}
\begin{equation}
\partial_{\theta_{23}} \tilde{H} = 
\frac{1}{2E}\left(
\partial_{\theta_{23}}U \cdot \text{diag}(0, \Delta m^2_{21}, \Delta m^2_{31}) \cdot U^\dagger + U \cdot \text{diag}(0, \Delta m^2_{21}, \Delta m^2_{31}) \cdot \partial_{\theta_{23}}U^\dagger \right)
\end{equation}

\subsection{\texorpdfstring{$\theta_{12}$}{th12}}

\begin{equation}
\partial_{\theta_{12}}U = \begin{pmatrix}
-U_{01} & U_{00} & 0 \\
-U_{11} & U_{10} & 0 \\
-U_{21} & U_{20} & 0
\end{pmatrix} \nonumber
\end{equation}
\begin{equation}
\partial_{\theta_{12}} \tilde{H} = 
\frac{1}{2E}\left(
\partial_{\theta_{12}}U \cdot \text{diag}(0, \Delta m^2_{21}, \Delta m^2_{31}) \cdot U^\dagger + U \cdot \text{diag}(0, \Delta m^2_{21}, \Delta m^2_{31}) \cdot \partial_{\theta_{12}}U^\dagger \right)
\end{equation}

\subsection{\texorpdfstring{$\theta_{13}$}{th13}}

\begin{equation}
\partial_{\theta_{13}}U = \begin{pmatrix}
-c_{12} s_{13} & -s_{12} s_{13} & c_{13} e^{i\delta} \\
-c_{12} s_{23} c_{13} e^{-i\delta_{CP}} & -s_{12} s_{23} c_{13} e^{-i\delta_{CP}} & -s_{23} s_{13} \\
-c_{12} c_{23} c_{13} e^{-i\delta_{CP}} & -s_{12} c_{23} c_{13} e^{-i\delta_{CP}} & -c_{23} s_{13}
\end{pmatrix} \nonumber
\end{equation}
\begin{equation}
\partial_{\theta_{13}} \tilde{H} = 
\frac{1}{2E}\left(
\partial_{\theta_{13}}U \cdot \text{diag}(0, \delta_{CP} m^2_{21}, \Delta m^2_{31}) \cdot U^\dagger + U \cdot \text{diag}(0, \Delta m^2_{21}, \Delta m^2_{31}) \cdot \partial_{\theta_{13}}U^\dagger \right)
\end{equation}

\subsection{\texorpdfstring{$\delta_{CP}$}{dcp}}

\begin{equation}
\partial_{\delta_{CP}}U = \begin{pmatrix}
0 & 0 & -i \, s_{13} e^{i\delta_{CP}} \\
c_{12} s_{23} s_{13} e^{-i\delta_{CP}} & s_{12} s_{23} s_{13} e^{-i\delta_{CP}} & 0 \\
c_{12} c_{23} s_{13} e^{-i\delta_{CP}} & s_{12} c_{23} s_{13} e^{-i\delta_{CP}} & 0
\end{pmatrix} \nonumber
\end{equation}
\begin{equation}
\partial_{\delta_{CP}} \tilde{H} = 
\frac{1}{2E}\left(
\partial_{\delta_{CP}}U \cdot \text{diag}(0, \Delta m^2_{21}, \Delta m^2_{31}) \cdot U^\dagger + U \cdot \text{diag}(0, \Delta m^2_{21}, \Delta m^2_{31}) \cdot \partial_{\delta_{CP}}U^\dagger \right)
\end{equation}

\subsection{Neutrino energy, \texorpdfstring{$E$}{ee}}

\begin{equation}
    \partial_{E} \tilde{H} = -\frac{1}{2E^2}\,
    U \cdot \text{diag}(0, \Delta m^2_{21}, \Delta m^2_{31}) \cdot U^\dagger +
    \frac{\Delta m^2_{21} + \Delta m^2_{31}}{6E^2}\,
\end{equation}

\subsection{Baseline, \texorpdfstring{$x$}{ll}}
The case of the baseline is different and simpler than the rest and we can simply use Schr\"ondiger's equation in Equation~\ref{eq:sch} without the need of applying de results in Section~\ref{sec:diff}.
\begin{equation}
    \frac{\partial}{\partial x}\tilde{\Psi} = -i\tilde{H}\tilde{\Psi},
\end{equation}

\end{document}